\documentclass[a4paper,12pt,draftclsnofoot,onecolumn,twoside]{IEEEtran}

\usepackage[cmex10]{amsmath}
\usepackage{mathrsfs}
\usepackage{amssymb}
\usepackage{bm}
\usepackage{units}
\usepackage{graphicx}
\usepackage{float} 
\usepackage{subfigure}

\usepackage{tikz}
\usepackage{diagbox}
\usepackage{makecell}
\usepackage{colortbl}
\usepackage{threeparttable}

\usepackage{algorithm}
\usepackage{algorithmic}

\usepackage{cite}
\usepackage[CJKbookmarks=true,bookmarksnumbered=true,bookmarksopen=true,linkcolor=green,anchorcolor=green,citecolor=green]{hyperref}
\usepackage{setspace}
\usepackage{lineno}
\usepackage{authblk}

\newtheorem{thm}{Theorem}

\newtheorem{lem}{Lemma}
\newtheorem{cor}{Corollary}

\begin{document}


\title{What Should Future Wireless Network Architectures Be?}
\author{Lu Yang, 
	\IEEEmembership{MIEEE,}
	Ping Li, Miaomiao Dong, Bo Bai, 
	\IEEEmembership{SMIEEE,} 
	Dmitry Zaporozhets, Xiang Chen, 
	Wei Han, \IEEEmembership{MIEEE,}  
	and Baochun Li, \IEEEmembership{FIEEE} 
	\thanks{L. Yang, P. Li, M. Dong, B. Bai, X. Chen, and W. Han are with the Theory Lab, Huawei Technologies Co. Ltd., Hong Kong Science Park, Shatin, New Territories, Hong Kong
		(e-mail: \{yang.lu4, liping129, dong.miaomiao, baibo8, chenxiang73, harvey.hanwei\}@huawei.com). (The 
			corresponding author is B. Bai.)}
	\thanks{D. Zaporozhets is with the Department of Steklov Mathematical Institute, Russian Academy of Sciences, St. Petersburg, Russia (email: zap1979@gmail.com).} 
    \thanks{B. Li is with the Department of Electrical and Computer Engineering, University of Toronto, Toronto, Canada (email: bli@ece.toronto.edu).}}

 \maketitle

\begin{abstract}
The accelerated convergence of digital and real-world lifestyles has imposed unprecedented demands on today's wireless network architectures, as it is highly desirable for such architectures to support wireless devices everywhere with high capacity and minimal signaling overhead. Conventional architectures, such as cellular architectures, are not able to satisfy these requirements simultaneously, and are thus no longer suitable for the future era.
In this paper, we propose a capacity-centric (C$^2$) architecture for future wireless communication networks. 
It is designed based on the principles of maximizing the number of non-overlapping clusters with the average cluster capacity guaranteed to be higher than a certain threshold, and thus provides a flexible way to balance the capacity requirement against the signaling overhead. Our analytical results reveal that C$^2$ has superior generality, wherein both the cellular and the fully coordinated architectures can be viewed as its extreme cases. Simulation results show that the average capacity of C$^2$ is at least three times higher compared to that of the cellular architecture. More importantly, different from the widely adopted conventional wisdom that base-station distributions dominate architecture designs, we find that the C$^2$ architecture is independent of base-station distributions, and instead user-side information should be the focus in future wireless network architecture designs.
\end{abstract}


\newpage
\section{Introduction}
With 5G maturing as the global standard for wireless communications, we are not only experiencing an explosive increase of mobile data traffic, but also witnessing the fusion of real and digital worlds \cite{ITUwhitepaper,5GMoment1}. This phenomenon is anticipated to make our lifestyles more intelligent and automated in the coming era. Various types of intelligent services are emerging, such as autonomous vehicles, immersive media, smart homes, remote healthcare, and factory automation. In 2019, the concept of ubiquitous wireless intelligence was introduced; wireless connectivity as a part of a critical infrastructure will provide services for both human and non-human users everywhere seamlessly through smart devices and applications \cite{Oulu19,6GSummit}. Nevertheless, despite all the initiatives emerging around next-generation wireless networks, its fundamental architecture remains largely undefined.

Cellular networks, which have been commercialized till 5G, are the most classic architectures. 
A cellular network consists of multiple cells, with each cell corresponding to the coverage area of a base station (BS) \cite{Cellular}. 
Theoretically, we often use a simplified model of hexagonal cell with a BS in the middle to represent cellular networks \cite{D.Tse}. 
In practice, the locations of deployed BSs are irregular. The coverage map of a cellular network with randomly-located BSs is a Voronoi tessellation \cite{Cellular_voronoi, Cellular_yanglu, Cellular_tractable}, as illustrated in Table \ref{Table:architecture_comparison}. A cellular architecture is BS-centric, and its main drawback is its poor signal quality at cell edges, where useful signals are affected by heavy interference from  other cells. 
To mitigate the cell-edge problem, an improved architecture called Coordinated Multi-Point transmission (CoMP) was proposed \cite{CoMP, DaiBai17}. With CoMP, several closely-located BSs are grouped as a cluster to serve users in a coordinated fashion, illustrated in Table \ref{Table:architecture_comparison}. 
A user moving inside a cluster enjoys continuous service, with the cell-edge problem avoided. However, heavy interference still exists at the cluster edges, since CoMP is BS-centric as well. As a result, neither the cellular architecture nor the CoMP architecture is suitable for future wireless communication networks, given that they cannot guarantee wireless services good enough for users densely located everywhere.

An alternative architecture designed to eliminate the cell-edge problem is a fully coordinated network, where all the BSs are inter-connected for information to be exchanged and optimal decisions to be made \cite{Coordinated,Interdonato19}. Unfortunately, though such an architecture works in theory, it is not feasible in real-world deployments due to its lack of scalability. In a fully coordinated network, all BSs take part in serving all users with one or several central processing units \cite{6Gchallenge}. The volume of information to be exchanged over fronthaul/backhaul links \cite{fronthaul, fronthaul_backhaul, user_centric_CRAN} expands significantly faster than the increase in the number of network nodes (BSs and users). 
A substantial amount of signaling overhead and time cost on information delivery, processing, and decision-making will be provoked. 
This is the reason why the number of inter-connected BSs is always limited in practice. At the late stages of 5G development, a new architecture called cell-free massive MIMO (multiple-input and multiple-output) was proposed and quickly became an active research topic \cite{Tariq19, Junyuan21, JunyuanICC}. The basic idea of being ``cell free'' is to eliminate cell boundaries through interconnecting all the massive MIMO antennas distributed in a geographic area to serve users coherently \cite{Ngo17}. There are no cells and thus no cell-edge problems \cite{6Gchallenge, UCNC_huawei}. However, the essence of such an architecture goes back to the design of a fully coordinated network. Although some ideas have been proposed to partially mitigate the lack of scalability \cite{Interdonato19ICC}, such an architecture is still far from being deployed in practice.

\begin{table}[H]
	\caption{Comparison of four wireless network architectures}
	\centering
	\setlength \tabcolsep{2pt}
	\label{Table:architecture_comparison}
	\small
	\begin{threeparttable}
	\begin{tabular}{ | l | c | c | c | c| } 
		\hline
		\textbf{Architecture} & \textbf{Cellular} & \textbf{CoMP} & \textbf{Fully Coordinated } & \textbf{Capacity Centric (C$^2$)} \\ 
		\hline \hline
		\makecell[l]{Toy example \\ of the coverage \\ map}
		&
		\makecell[c]{
			\begin{minipage}[b]{0.175\columnwidth}
				\centering
				\raisebox{-.5\height}{\includegraphics[width=\linewidth]{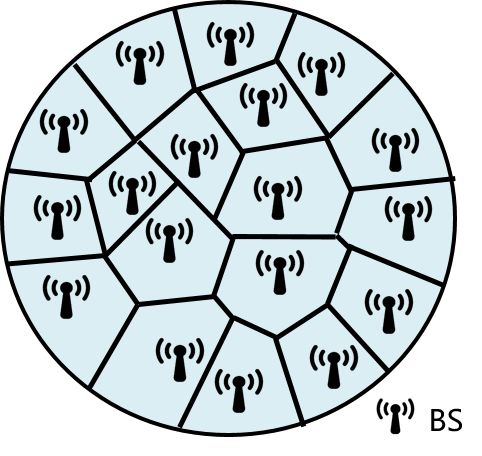}}
			\end{minipage}}
		& 
		\makecell[c]{
			\begin{minipage}[b]{0.17\columnwidth}
				\centering
				\raisebox{-.5\height}{\includegraphics[width=\linewidth]{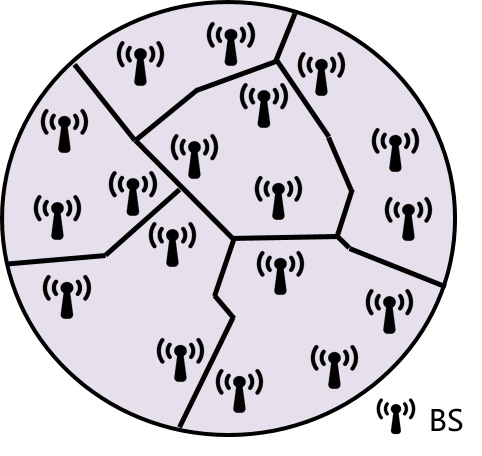}}
			\end{minipage}}
		&
		\makecell[c]{
			\begin{minipage}[b]{0.18\columnwidth}
				\centering
				\raisebox{-.5\height}{\includegraphics[width=\linewidth]{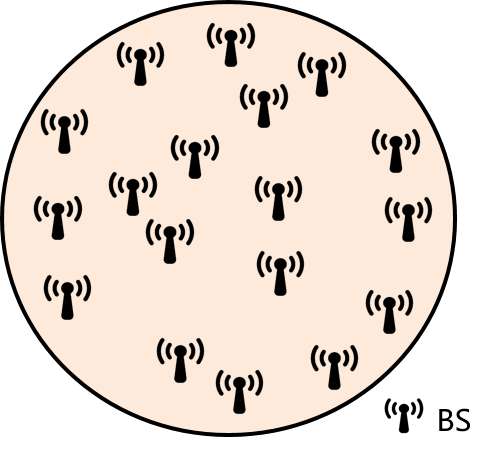}}
			\end{minipage}}
		&
		\makecell[c]{
			\begin{minipage}[b]{0.2\columnwidth}
				\centering
				\raisebox{-.5\height}{\includegraphics[width=\linewidth]{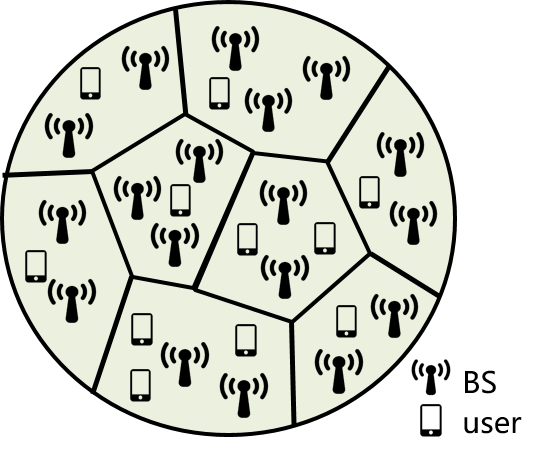}}
			\end{minipage}}
		\\ \hline
		\makecell[l]{Design \\ principle}  
		& \makecell[l]{Coverage area of a \\ single BS is a cluster  }
		& \makecell[l]{Several closely- \\ located BSs form \\ a cluster }    
		& \makecell[l]{All BSs participated \\ in serving all users }  
		& \makecell[l]{Average cluster capacity \\ guaranteed and signaling \\overhead minimized}    
		\\ \hline
				\makecell[l]{Architecture\\ type}  
		& \multicolumn{3}{c|}{\makecell[c]{BS centric}} 
		& \makecell[c]{Capacity centric}    
		\\ \hline
			    \makecell[l]{Architecture \\ robustness}
		& \multicolumn{3}{c|}{\makecell[c]{Low \\ The architecture changes with different BS distributions}}
		& \makecell[c]{High } 
		\\ 
		\hline
				Tessellation 
		& \makecell[c]{Voronoi Tessellation} 
		& \makecell[c]{BS-centric \\ Tessellation} 
		& -- 
		& \makecell[c]{C$^2$ Tessellation  \\ (C$^2$T)} 
		\\ 
		\hline
		\makecell[l]{Relationship \\ with C$^2$T}
		& \makecell[c]{Extreme case of C$^2$T \\ ($M=L$)\tnote{$\dag$}} 
		& --
		& \makecell[c]{Extreme case of C$^2$T \\ ($M=1$)} 
		& \makecell[c]{C$^2$T} 
		\\ \hline

		Scalability
		& High & Medium & Low & High
		\\ \hline
		\makecell[l]{Signaling \\ overhead}
		& Low & Medium & High & Low
		\\ \hline
		\makecell[l]{Capacity} & 
		\multicolumn{4}{c|}{ The average cluster capacity per BS can be determined according to Theorem \ref{thm:Cauchy}.}
		\\ \hline
	\end{tabular} 
	   \begin{tablenotes} 
	    \item [$\dag$] $M$ denotes the number of clusters. $L$ denotes the number of BSs.
	   \end{tablenotes} 
    \end{threeparttable}
\end{table}

The major contributions of this paper are summarized as follows.
\begin{enumerate}
	\item A Capacity-Centric (C$^2$) architecture for future wireless communication networks is proposed. 
	It decomposes the entire network into non-overlapping clusters, with the objective of maximizing the number of clusters and guaranteeing the average cluster capacity per BS (or per user) larger than a predefined threshold.
	On one hand, by guaranteeing the average cluster capacity per BS (or per user) larger than a predefined threshold, all the users can enjoy good-enough wireless services, and the edge problem in BS-centric architectures is therefore solved in C$^2$. 
	On the other hand, by maximizing the number of non-overlapping clusters, the network possesses high scalability with the signaling overhead kept at a minimal level. This is because only BSs belonging to the same cluster are coordinated, and adding or removing network nodes in a cluster will not affect the signaling overhead of other clusters. The unscalability problem in fully coordinated architectures can be solved in C$^2$ as well.
	Thus, the C$^2$ architecture illustrates its potential to improve the overall network performance, while keeps the signaling overhead and computation complexity under control.
	
	\item  A general and simplified method to determine the capacity is proposed.
	Specifically, a novel expression of the average cluster capacity per BS (or per user) under the assumption that the numbers of BSs and users approach infinity is derived, as detailed later in Section II-C. 
	This expression is one of the most important foundations of our C$^2$ architecture, and is also widely applicable for other architectures with different BS and user distributions. Based on this new expression, the capacity region and related performance metrics of an  ultra-dense network can be derived with a much lower complexity.
	
	\item 
	A comprehensive comparison of different network architectures is presented, and the superiority of our C$^2$ architecture is shown through performance evaluation. 
    Table \ref{Table:architecture_comparison} summarizes the characteristics of four different architectures: cellular, CoMP, fully coordinated and C$^2$. 
    Our C$^2$ architecture design takes the information of both BSs and users into consideration, since the capacity is determined by BSs and users jointly.
	This is the major difference between C$^2$ and conventional BS-centric architectures (cellular and CoMP), since the latter is designed to focus on the BS-side information only.
	In Section III, the generality of the C$^2$ architecture is presented. Specifically, C$^2$ can cover almost all the network configurations, wherein the cellular and the fully coordinated architectures can be regarded as two extreme cases. Our simulation results in Section V show that the C$^2$ architecture outperforms cellular and CoMP networks, with performance gains of at least $300\%$ and $17\%$, respectively. More importantly, both our theoretical and simulation results disclose that the user distribution plays a non-negligible role in network design, while the BS distribution will not materially affect the network architecture. For this reason, currently commercialized BS-centric designs ignoring the user distribution is no longer suitable for future wireless communication networks.
	
\end{enumerate}

The following content of the paper is organized as follows.
The basics of C$^2$ architecture designs, including the system model, the problem formulation, and the average capacity theorem are introduced in Section II.
Section III elaborates the procedures to determine the C$^2$ architectures with specific network settings, where we take the setting of a constant user density as an example.
Section IV visualizes the C$^2$ architectures based on theoretical results, and derives the capacity region of C$^2$.
Simulation results on performance comparison across C$^2$, CoMP, and cellular architectures are shown in Section V. 
Section VI concludes the paper, and some detailed proofs are relegated to appendices.

\textit{Notation:}
In this paper, scalars and matrices are represented by lowercase and bold uppercase letters, like $h$ and $\mathbf{H}$, respectively, and the $(i,j)$-th entry of $\mathbf{H}$ is denoted by $[\mathbf{H}]_{i,j}$.
Bold lowercase letters $\textbf{x}$ and $\textbf{y}$ denotes the location coordinates of the BSs and the users, respectively. 
Other bold lowercase letters, like $\mathbf{h}$, represents the vectors.
Scripts such as $\mathcal{B}$ denote the sets. 
Notation $\mathcal{CN}(\mu,\Gamma)$ refers to a complex normal distribution with mean $\mu$ and variance $\Gamma$.
Operators $(.)^H$, $\det(.)$ and $\mathbb{E}\{.\}$ represent the Hermitian transpose, determinant and expectation, respectively. 

\section{C$^2$ design principles and average capacity theorem}

\subsection{System Model}
Consider a wireless communication network that constitutes two kinds of nodes: $L$ single-antenna BSs (or access points in distributed-antenna systems \cite{Buzzi17,Dai_DAS,singleAntenna,Alex}) and $K$ single-antenna users \cite{singleAntenna, single_antenna}. 
Denote the set of BSs as $\mathcal{B} = \left\{ b_1, b_2, ..., b_L \right\}$, and the set of users as $\mathcal{U} = \left\{ u_1, u_2, ..., u_K \right\}$. 
For a cluster with index $j$ in this network, 
we use $\mathcal{C}_j$ to denote the union set of BSs and users in it.
The number of BSs in $\mathcal{C}_j$ is denoted by $L_j$, and the number of users in $\mathcal{C}_j$ is denoted by $K_j$.

The channel gain between BS $b_l \in \mathcal{C}_j$ and user $u_k\in\mathcal{U}$ is $h_{jlk}$, which can be calculated according to
\begin{equation}
h_{jlk}= \ell_{jlk} \times g_{jlk}.
\label{eq:h_jlk}
\end{equation}
Here,
$g_{jlk}\sim\mathcal{CN}(0,1)$ is the small-scale fading.  $\ell_{jlk}$ is the large-scale fading, which describes the signal attenuation between BS $b_l$ and user $u_k$ as a function of the signal propagation distance \cite{Buzzi18},  defined as
\begin{equation}
\ell_{jlk} = \left( \theta d_{jlk}^{-\alpha} \right) ^{1/2},
\label{eq:path-loss}
\end{equation}
where $d_{jlk}$ is the Euclidean distance between $b_l$ and $u_k$. Parameters $\theta$ and $\alpha$ are constants and can take different values based on different path-loss models. 
In this paper, we use the general form of $\ell_{jlk}$ \eqref{eq:path-loss}  for theoretical analysis, and choose an example path-loss model with specific values of $\theta$ and $\alpha$ \cite{Buzzi18} for simulation in Section IV.

The uplink signal model of cluster $j$ is given by
\begin{equation}
\mathbf{y}_j=\sum_{u_k\in\mathcal{C}_j}\mathbf{h}_{jk}s_k+
\sum_{u_\kappa\in {\mathcal{U}\setminus\mathcal{C}_j} }\mathbf{h}_{j\kappa}s_\kappa+\mathbf{z}_j,
\end{equation}
where $\mathbf{h}_{jk}$ is an $L_j \times 1$ vector, whose $l$-th entry is $h_{jlk}$. 
 $s_k\sim\mathcal{CN}(0,P)$ is the information-bearing signal of the user $u_k$, with $P$ as the transmit power of each user.
  $\mathbf{z}_j\sim\mathcal{CN}(\mathbf{0},N_0\mathbf{I})$ is the additive white Gaussian noise (AWGN) vector.

The average capacity per BS of cluster $j$ is  \cite{D.Tse,Shannon}
\begin{equation}
C_j=\mathbb{E}\left\{\frac{1}{L_j}\log\det\left(\mathbf{I}+ P (N_0\mathbf{I}+ 
\sum_{u_\kappa \in {\mathcal{U}\setminus\mathcal{C}_j}} \mathbf{h}_{j\kappa}\mathbf{h}_{j\kappa}^H)^{-1}\mathbf{H}_j  \mathbf{H}_j^H\right)\right\},
\end{equation}
where $\mathbf{H}_j$ is the channel gain matrix of cluster $j$, with $[\mathbf{H}_j]_{l,k}=h_{jlk}$.
This expression is equivalent to
\begin{equation}
C_j=\mathbb{E}\left\{\frac{1}{L_j} \log \det\left(\mathbf{I}+ P (N_0\mathbf{I}+\mathbf{\Sigma}_j)^{-\frac{1}{2}}\mathbf{H}_j \mathbf{H}_j^H(N_0\mathbf{I}+\mathbf{\Sigma}_j)^{-\frac{1}{2}}\right)\right\},
\label{eq:capacity_Cj}
\end{equation}
where $\mathbf{\Sigma}_j$ denotes the inter-cluster interference, and can be regarded as a diagonal matrix with the $l$-th diagonal entry as
\begin{equation}
[\mathbf{\Sigma}_j ]_{l,l}
=\sum_{{u_\kappa\in\mathcal{U}\setminus\mathcal{C}_j}}\mathbb{E}\left\{|h_{jl\kappa}|^2\right\}{P}
= {P} \sum_{u_\kappa\in\mathcal{U}\setminus\mathcal{C}_j} \ell_{jl\kappa}^{2}
= {P} \sum_{u_\kappa\in\mathcal{U}\setminus\mathcal{C}_j} \theta d_{jl\kappa}^{-\alpha}.
\end{equation}
If we define the equivalent channel gain matrix $\tilde{\mathbf{H}}_j$ as
\begin{equation}
\tilde{\mathbf{H}}_j=(N_0\mathbf{I}+\mathbf{\Sigma}_j)^{-\frac{1}{2}}\mathbf{H}_j.
\label{eq:H_j_tilde}
\end{equation}
The average cluster capacity per BS \eqref{eq:capacity_Cj} becomes
\begin{equation}
C_j = \mathbb{E} \left[ \frac{1}{L_j} \log \det\left(\mathbf{I}+ P \tilde{\mathbf{H}}_j  \tilde{\mathbf{H}}_j^H\right) \right].
\end{equation}

To capture the ultra-dense characteristics of future networks, we assume that both $K$ and $L$ approach infinity, and the ratio between $K$ and $L$ is denoted as $\beta$.
Similarly, for the cluster $j$, we assume both $K_j$ and $L_j$ approach infinity with  $\frac{K_j}{L_j} = \beta_j$. 
Denote $C_j^{\infty}$ as the average cluster capacity per BS under such asymptotic assumptions, and we have
\begin{equation}
 C_j^\infty 
= \lim_{K_j,L_j\to\infty}C_j 
 = \lim_{K_j,L_j\to\infty} \mathbb{E} \left[ \frac{1}{L_j} \log \det\left(\mathbf{I}+ P \tilde{\mathbf{H}}_j  \tilde{\mathbf{H}}_j^H\right) \right].
\label{eq:C_j_infty}
\end{equation}

\subsection{ Design Principles and Problem Formulation}
Our objective is to design a network architecture $\mathcal{M}$, which not only provides high-quality wireless services everywhere, but also possesses good scalablility suitable for real-world deployments.
Thus, our design principles of a brand-new architecture towards future wireless communication networks include the following two aspects. 
On one hand, to guarantee high-quality wireless services everywhere, we set the average cluster capacity larger than a predetermined threshold $C_{Th}$. This threshold can be designed based on specific requirements in practice. 
On the other hand, to achieve good scalability, we organize all the network nodes into $M$ ($ 1 \le M \le L$) disjoint clusters. BSs in the same cluster coordinate to serve the users, but work independently if located in different clusters. 
As such, information is exchanged inside each cluster. The scale of the induced signaling overhead is restricted by the cluster size. 
Different from a fully coordinated network,
the signaling overhead of the entire network will not fluctuate substantially due to the increase or decrease of a single network node. 
More specifically, we should maximize the number of clusters $M$, since the larger $M$ corresponds to the smaller cluster size and the less signaling overhead.
In what follows, we will present a detailed theoretical study that incorporate these principles for designing our C$^2$ architecture.

First, the problem of maximizing the number of clusters while satisfying the capacity requirement can be formulated in a concise form as
\begin{equation}
\begin{aligned}
{\mathcal{P}1:} \quad  &  \max_{\mathcal{M}} \quad M , \\
& ~~ s.t. \quad  C_j ^ {\infty} \ge C_{Th}, ~~\forall j \in \{1,2,\ldots,M\}. \\
\end{aligned}
\label{eq:obj}
\end{equation}
Note that $C_j^{\infty}$ is the average cluster capacity per BS, and the average cluster capacity per user is $C_j^\infty/\beta_j$, which is a scaled version of $C_j^{\infty}$.
Thus, the constraint in $\mathcal{P}$1 guarantees the communication quality of each BS or each user larger than a predefined threshold. 
Next, to solve $\mathcal{P}$1, we start from analyzing $C_j^\infty$.

\subsection{Average Cluster Capacity Theorem}
It can be observed from \eqref{eq:C_j_infty} that high-dimensional matrix manipulation should be performed in order to determine $C_j^{\infty}$. 
This will lead to high difficulty and complexity to solve  $\mathcal{P}$1.
In the following, we will introduce our proposed theorem to determine $C_j^{\infty}$ in a much simplified and ingenious way.

Based on the concavity of the log-determinant function, we have
\begin{equation}
\begin{aligned}
C_j^\infty 
& = \lim_{K_j,L_j\to\infty} \mathbb{E} \left[ \frac{1}{L_j} \log \det\left(\mathbf{I}+ P \tilde{\mathbf{H}}_j  \tilde{\mathbf{H}}_j^H\right) \right] \\
& =
\lim_{K_j,L_j\to\infty}   \frac{1}{L_j}\sum_{l=1}^{L_j} \log \left(1+ P \lambda_{lj}\right),  
\end{aligned} 
\label{eq:C_j_infty_lambda}
\end{equation}
where  $\lambda_{lj}$ is the $l$th $(L=1,2,\ldots, L_j)$ eigenvalue of matrix $\tilde{\mathbf{H}}_j  \tilde{\mathbf{H}}_j^H$.
Note that under the asymptotic assumption of $L_j, K_j \to \infty$,
$\tilde{\mathbf{H}}_j\tilde{\mathbf{H}}_j^H $ converges to a diagonal matrix \cite{Chebyshev}, and thus its diagonal entries are its eigenvalues.
Then based on \eqref{eq:H_j_tilde} and \cite{Chebyshev},
the eigenvalues of $\tilde{\mathbf{H}}_j\tilde{\mathbf{H}}_j^H $ can be derived as
\begin{equation}
\lambda_{lj} 
= \frac{\sum_{u_k\in\mathcal{U}_j}\theta d_{jlk}^{-\alpha}}{N_0  +  \sum_{u_\kappa\in\mathcal{U}\setminus\mathcal{C}_j} \theta d_{jl\kappa}^{-\alpha} {P} } , \quad \forall ~ l \in \{1,2,...,L_j\}.
\label{eq:lambda_j_def}
\end{equation}
For further analysis, with the asymptotic assumptions,
the above expression (\ref{eq:lambda_j_def}) can be transformed by replacing the discrete distributions of the network nodes by continuous density as follows.

Let $\mathcal{D}_0$ denote the two-dimensional region spanning the entire network, and let $\mathcal{D}_j \subseteq \mathcal{D}_0$ denote the region spanned by the $j$-th cluster. 
Note that the set $\{\mathcal{D}_1, \mathcal{D}_2, \ldots, \mathcal{D}_M \}$ is a partition of $\mathcal{D}_0$.
With a slight abuse of notation, we use $|\mathcal{D}_0|$ to denote the area of  $\mathcal{D}_0$, and $|\mathcal{D}_j|$ to denote the area of $\mathcal{D}_j$.
Assume BSs and users are distributed over  $\mathcal{D}_0$ according to continuous density functions ${\rho_b}(\textbf{x})$ and $\rho_u(\textbf{y})$, with $\textbf{x}$ and $\textbf{y}$ representing the location coordinates of the BSs and users, respectively. 
Then $\lambda_{lj}$ \eqref{eq:lambda_j_def} can be represented by a continuous form as
\begin{equation}
\lambda_{lj} = \frac{ \int_{\textbf{y} \in \mathcal{D}_j} f(\textbf{x} - \textbf{y}) \rho_u (\textbf{y}) d\textbf{y}}{N_0  + P \int_{\textbf{y} \in \mathcal{D}_0 \setminus \mathcal{D}_j } f(\textbf{x} - \textbf{y}) \rho_u (\textbf{y}) d\textbf{y}} , \quad (l=1,2,...,L_j),
\end{equation}
where 
\begin{equation}
f(\textbf{x}-\textbf{y}) = \theta d_{\textbf{xy}}^{-\alpha},
\end{equation}
defined as the square of large-scale fading, with $d_{\textbf{xy}}$ denoting the Euclidean distance between the BS with coordinates $\textbf{x}$ and the user with coordinates $\textbf{y}$.
Similarly, by transforming other items in \eqref{eq:C_j_infty_lambda} from discrete forms into continuous ones, and based on Cauchy's Mean Value Theorem \cite{CauchyMean},
 we derive a new expression to determine ${C}_j^{\infty}$ as follows.

\smallskip
\begin{thm}[Average cluster capacity]
	\label{thm:Cauchy} 
	Conditioning on the number of BSs and the number of users approaching infinity,
	the average cluster capacity per BS is given by
	\begin{equation}
	C_j^{\infty}
	= 
	{\log}_2 ~\frac{ \frac{N_0}{P} + \int_{\textbf{y} \in \mathcal{D}_0} f(\textbf{x}_{jl} - \textbf{y}) \rho_u (\textbf{y}) d\textbf{y}}
	{ \frac{N_0}{P} + \int_{\textbf{y} \in \mathcal{D}_0 \setminus \mathcal{D}_j} f(\textbf{x}_{jl} - \textbf{y}) \rho_u(\textbf{y}) d\textbf{y}} ,
	\label{eq:Cj_general}
	\end{equation}
	for some BS $b_l$ in the $j$th cluster with coordinate denoted by $\textbf{x}_{jl}$. 

\begin{IEEEproof}
	By transforming all the items in \eqref{eq:C_j_infty_lambda} from discrete forms into continuous ones, 
	including replacing the discrete distributions of the network nodes by continuous distributions, and replacing the summation by the integral,
	we have
	\begin{equation}
	\begin{aligned}
	C_j^\infty &= \lim_{K_j,L_j\to\infty}\frac{1}{L_j}\sum_{l=1}^{L_j}\log_2\left(1+ {P} \lambda_{lj}\right)\\
	& = \lim_{K_j,L_j\to\infty}\frac{1}{L_j}\sum_{l=1}^{L_j}\log_2\left(\frac{\frac{N_0}{P} +\sum_{u_k\in\mathcal{U}}  \ell_{jlk}^{2}}{\frac{N_0}{P}+\sum_{u_k\in\mathcal{U}\setminus\mathcal{C}_j} \ell_{jlk}^{2}}\right)\\
	&= \frac{1}{\int_{ \textbf{x} \in \mathcal{D}_j} \rho_b(\textbf{x}) d\textbf{x}}  \int_{\textbf{x} \in \mathcal{D}_j} \rho_b(\textbf{x}) \log_2 \frac{\frac{N_0}{P} + \int_{\textbf{y} \in \mathcal{D}_0} f(\textbf{x}-\textbf{y}) \rho_u(\textbf{y}) d\textbf{y}} {\frac{N_0}{P} + \int_{\textbf{y} \in \mathcal{D}_0 \setminus \mathcal{D}_j} f(\textbf{x}-\textbf{y}) \rho_u (\textbf{y}) d\textbf{y}} d\textbf{x}\\
	& = \log_2 \frac{\frac{N_0}{P} + \int_{\textbf{y} \in \mathcal{D}_0} f(\textbf{x}_{jl} - \textbf{y}) \rho_u (\textbf{y}) d\textbf{y}}{\frac{N_0}{P} + \int_{\textbf{y} \in \mathcal{D}_0 \setminus \mathcal{D}_j} f(\textbf{x}_{jl} - \textbf{y}) \rho_u(\textbf{y}) d\textbf{y}},
	\end{aligned}
	\end{equation}
	for some $\textbf{x}_{jl} \in \mathcal{D}_j$, where the last equation follows Cauchy's Mean Value Theorem \cite{CauchyMean}. \\
	The proof is completed.
\end{IEEEproof}
\end{thm}

\medskip

Theorem \ref{thm:Cauchy} provides a substantially simplified expression to determine $C_j^{\infty}$, where neither the high-dimensional matrix manipulation nor the eigenvalue derivation is needed. 
This expression is sufficiently general to be applied in various networks, since it does not have any constraints on the network node distributions, the network area, and the cluster areas.
In addition, Theorem \ref{thm:Cauchy} reveals that the user density function $\rho_u(\textbf{y})$ plays an important role in determining $C_j^{\infty}$, and thus further affects the network architecture design $\mathcal{M}$. 
Conventional BS-centric architectures have their disadvantages such as cell-edge problems, due to the missing role from the user's side.

\section{The C$^2$ architecture with a constant user density}
In this section, we will elaborate how a C$^2$ architecture is to be derived with a specific network setting. 
As aforementioned, both the problem formulation $\mathcal{P}1$ and Theorem \ref{thm:Cauchy} are adaptive to different network settings. 
In the following, we will take the setting of a constant user density as an example case to derive the C$^2$ architecture. 

With a constant user density, $\rho_u(\textbf{y})$ can be simplified as $\rho_u$, unrelated to user location coordinates $\textbf{y}$. 
Such a distribution is also known as a Poisson distribution. 
By substituting $\rho_u$ for $\rho_u(\textbf{y})$ in (\ref{eq:Cj_general}), and focusing on the interference-limited regime where the background noise can be ignored, we can determine the average cluster capacity per BS according to the following corollary.
\smallskip
\begin{cor}[Average cluster capacity with a constant user density]
	\label{cor:Cauchy_uniform}	  
   Conditioning on a constant user density, and the number of BSs and users approaching infinity, the average cluster capacity per BS is given by
	\begin{equation}
	C_j^{\infty}
	 =  {\log}_2
	\frac{\int_{\textbf{y} \in \mathcal{D}_0} f(\textbf{x}_{jl} - \textbf{y}) d\textbf{y}}{\int_{\textbf{y} \in \mathcal{D}_0 \setminus \mathcal{D}_j} f(\textbf{x}_{jl} - \textbf{y}) d\textbf{y}},
	\label{eq:Cauchy_uniform}
	\end{equation}
	for some BS $b_l$ in the $j$th cluster with coordinates denoted by $\textbf{x}_{jl}$.  
\end{cor}
\smallskip

Corollary \ref{cor:Cauchy_uniform} reveals that with a constant user density and a given cluster region, $C_j^{\infty}$ is determined by a certain BS location inside this cluster.
In addition, it reveals that
$C_j^{\infty}$ is positively correlated to the cluster area $\left|\mathcal{D}_j\right|$. As such, the constraints in $\mathcal{P}$1 can be transformed to the cluster area $\left|\mathcal{D}_j\right|$ larger than a given area threshold $\left|\mathcal{D}_{Th}\right|$. The problem of maximizing the number of clusters while satisfying the capacity requirement under the condition of a constant user density can be reformulated as
\begin{equation}
\begin{aligned}
{\mathcal{P}2:} \quad  &  \max_{\mathcal{M}} \quad M , \\
& ~~ s.t. \quad  \left|\mathcal{D}_j\right|  \ge \left|\mathcal{D}_{Th}\right|, ~~\forall j \in \{1,2,\ldots,M\}.
\end{aligned}
\label{eq:obj_uniform_distribution}
\end{equation}
Combining the constraint in $\mathcal{P}2$ and the fact of $\sum_{j=1}^M \left|\mathcal{D}_j\right| = |\mathcal{D}_0|$, we can obtain the optimal number of clusters $M^*$ and the optimal cluster area $|\mathcal{D}_j^*|$ as
\begin{equation}
M^* = \left\lfloor \frac{|\mathcal{D}_0|}{|\mathcal{D}_{Th}|} \right\rfloor,~~ 
\vert\mathcal{D}_j^*\rvert = \vert\mathcal{D}_{Th}\rvert,
\label{eq:M*Dj*}
\end{equation}
where $\lfloor x \rfloor$ is the floor function giving the greatest integer less than or equal to the input value $x$. 
We can conclude that an optimal network architecture keeps the signaling overhead at a minimal level while guaranteeing good-enough wireless services, and can be designed based on (\ref{eq:M*Dj*}).

Let's go a step further to analyze the optimal solutions (\ref{eq:M*Dj*}). If we set $\vert \mathcal{D}_{Th} \rvert$ to be equal to $|\mathcal{D}_0|$, the extreme case $M^*=1$ arises, corresponding to a fully coordinated network architecture, with a maximized network capacity but the highest signaling overhead. If $|\mathcal{D}_{Th}| = |\mathcal{D}_0|/{L}$, another extreme case $M^*=L$ arises, corresponding to the cellular architecture, with the number of clusters equal to the number of BSs. The cluster capacity in this case is the worst, but the signaling overhead of each cluster is the lowest. As such, we can claim that our C$^2$ architecture is sufficiently general, in which both the cellular and fully coordinated networks can be regarded as extreme cases. 

Another important finding is that our derived results for network design are independent of BS distributions. As long as users are distributed with a constant density, the C$^2$ architecture can be determined directly according to (\ref{eq:M*Dj*}).
BS distributions should not be regarded as the dominating factor in future architecture designs. In other words, next-generation wireless network architectures should not be BS-centric.

\section{C$^2$ visualization and capacity region analysis}
Based on our derived results in Section III, we now visualize the C$^2$ architectures and perform the capacity analysis with more explicit network settings, including the path-loss parameters, the shapes of the network region $\mathcal{D}_0$ and the cluster region $\mathcal{D}_j$. 

\subsection{Visualization of $C^2$ Architectures}
Assume both $\mathcal{D}_0$ and $\mathcal{D}_j$ are round with radius denoted by $R_0$ and $R_j$, respectively. 
The centers of $\mathcal{D}_0$ and $\mathcal{D}_j$ are denoted as $O_0$ and $O_j$, with coordinates $\textbf{x}_{O_0}$ and $\textbf{x}_{O_j}$, respectively.
With a constant user density, the optimal number of clusters $M^*$ and the optimal cluster radius $R_j^*$ for all $j \in \{1,2,\ldots,M\}$ can be derived based on  (\ref{eq:M*Dj*}) as
 \begin{equation}
M^* = \left\lfloor \frac{R_0^2}{R_{Th}^2} \right\rfloor,~~ 
R_j^* = R_{Th}.
\label{eq:M*Rj*}
\end{equation}
Note that these results are obtained without constraints on the distribution of BSs.

\begin{figure}[htbp]
	\centering
	\subfigure[]
	{\includegraphics[width=0.48\textwidth]{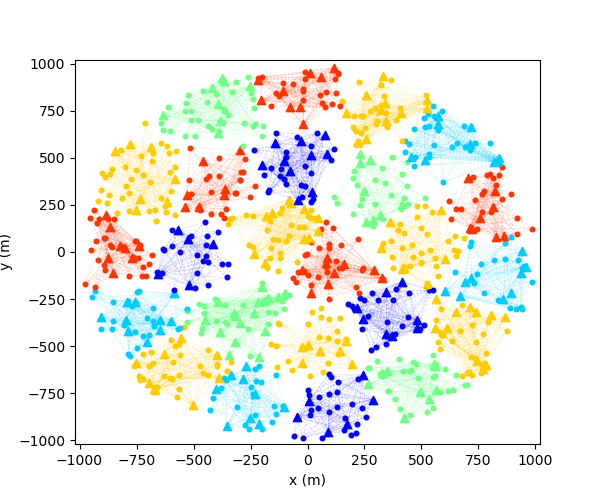}
		\label{fig:AUNT_uniformBS}} 	
	~\subfigure[]
	{\includegraphics[width=0.48\textwidth]{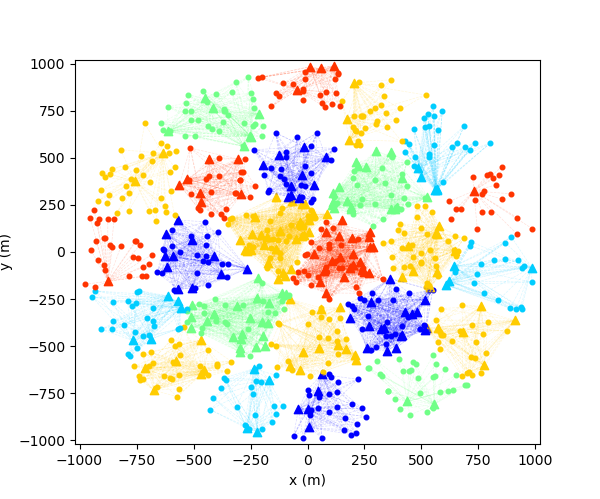}
		\label{fig:AUNT_nonuniformBS}} 	
	\caption{\textbf{Visualization of C$^2$ architectures with a constant user density but different BS densities.}
		Triangles are BSs. Circles are users.
		Each cluster is presented as a group of inter-connected nodes by lines in the same color.
		\textbf{(a)} C$^2$ architecture with a constant BS density. 
		\textbf{(b)} C$^2$ architecture with an inconstant BS density. BSs are centrally dense and peripherally sparse.}
	\label{fig:demo}
\end{figure}

For visualization, we plot two C$^2$ architectures in Fig. \ref{fig:demo}. 
The case of BSs deployed with a constant density $\rho_u$ is shown in Fig. \ref{fig:AUNT_uniformBS}. 
The case of BSs deployed centrally dense and peripherally sparse 
is shown in Fig. \ref{fig:AUNT_nonuniformBS}. 
Specifically, to simulate the latter case,  we divide the entire round network region $\mathcal{D}_0$ into three concentric subregions, with radius intervals $\left[0, \frac{1}{3} R_0 \right)$, $\left[\frac{1}{3} R_0, \frac{2}{3} R_0 \right)$, and $\left[\frac{2}{3} R_0, R_0 \right]$, respectively.
With a given $\rho_u$, $\rho_{b}$ can be calculated according to $\rho_{b} = \rho_u/\beta$. Then, we choose three different BS densities: $\frac{5}{9} {\rho_b}$, $\frac{3}{9} {\rho_b}$, $\frac{1}{9} {\rho_b}$, and allocate them to the three concentric subregions from the innermost to the outermost 
sequentially.  
Such a setting is to emulate real-world scenarios with a larger number of BSs in the central urban area, and fewer BSs in the surrounding rural area.
It can be observed from Fig. \ref{fig:demo} that both $M^*$ and $R_j^*$ are the same for the above two cases. It implies that the C$^2$ architecture is unrelated to the BS distributions, as long as the users are distributed with a constant density.

\subsection{Cluster Capacity Region of C$^2$ Architectures}
The capacity region is one of the most important performance metrics for wireless networks, but its derivation is always complex and difficult. 
In this subsection, we will study the minimum and maximum values for $C_j^{\infty}$.
Denote the maximum value of $ {C_j^{\infty}}$ as $\left. {C_j^{\infty}} \right|_{\max}$, and the minimum value of $ {C_j^{\infty}}$ as  $\left. {C_j^{\infty}} \right|_{\min}$. 
To derive the expressions of $\left. {C_j^{\infty}} \right|_{\max}$ and  $\left. {C_j^{\infty}} \right|_{\min}$, concrete path-loss model should be given. 
In this paper, we consider a three-slope model \cite{Buzzi18}, with path-loss parameters $\theta$ and $\alpha$ defined as  
\begin{equation}
\theta  = \left\{ {\begin{array}{*{20}{c}}
	1, \hfill & {{d_{\textbf{xy}}} > {d_1},} \hfill  \\
	{d_1^{ - 1.5}}, \hfill & {{d_0} < {d_{\textbf{xy}}} \le {d_1},} \hfill  \\
	{d_1^{ - 1.5}d_0^{ - 2}}, \hfill & { 0< {d_{\textbf{xy}}} \le {d_0},} \hfill  \\
	\end{array}} \right.
\quad
\alpha  = \left\{ {\begin{array}{*{20}{c}}
	{3.5}, \hfill & {{d_{\textbf{xy}}} > {d_1},} \hfill  \\
	2, \hfill & {{d_0} < {d_{\textbf{xy}}} \le {d_1},} \hfill  \\
	0, \hfill & { 0< {d_{\textbf{xy}}} \le {d_0},} \hfill  \\
	\end{array}} \right.
\label{eq:theta&alpha}
\end{equation}
where $d_0$ and $d_1$ can be interpreted as the near-field boundary and the far-field boundary, respectively.

According to our Corollary \ref{cor:Cauchy_uniform},  both $\left. {C_j^{\infty}} \right|_{\max}$ and  $\left. {C_j^{\infty}} \right|_{\min}$ can be determined if we find their corresponding BS locations.
The process to determine $\left. {C_j^{\infty}} \right|_{\max}$ and  $\left. {C_j^{\infty}} \right|_{\min}$ are elaborated as follows.
Based on \eqref{eq:C_j_infty_lambda}, 
we
 define
 \begin{equation}
 C_j^{\infty}  = \log {\Lambda_j},
 \label{eq:Cj_infty_&_Lambda_j}
 \end{equation}
 where
\begin{equation}
{\Lambda _j} = \lim_{K_j,L_j\to\infty} {\left[ {\prod\limits_{l = 1}^{{L_j}} {\left( {1 + {P}{\lambda _{lj}}} \right)} } \right]^{\frac{1}{L_j}}}.
\label{eq:Lambda_j_def}
\end{equation}
To determine the range of $C_j^{\infty}$, we start from analyzing $\Lambda_j$. 
Based on Corollary 1 and \eqref{eq:Cj_infty_&_Lambda_j}, we have
\begin{equation}
\frac{1}{\Lambda_j} 
= \frac{\int_{\textbf{y} \in \mathcal{D}_0 \setminus \mathcal{D}_j} f(\textbf{x}_{jl} - \textbf{y}) d\textbf{y}}{\int_{\textbf{y} \in \mathcal{D}_0} f(\textbf{x}_{jl} - \textbf{y}) d\textbf{y}}
= 1 - \frac{\int_{\textbf{y} \in  \mathcal{D}_j} f(\textbf{x}_{jl} - \textbf{y}) d\textbf{y}}{\int_{\textbf{y} \in \mathcal{D}_0} f(\textbf{x}_{jl} - \textbf{y}) d\textbf{y}},
\label{eq:1/Lambda_j_cauchy}
\end{equation}
for some BS $b_l$ with coordinates $\textbf{x}_{jl} \in \mathcal{D}_j$.
Then we define a function for all the BSs in $\mathcal{D}_j$, as
\begin{equation}
V_j(\textbf{x})
= 1 - \frac{\int_{\textbf{y} \in  \mathcal{D}_j} f(\textbf{x} - \textbf{y})   	   d\textbf{y}}{\int_{\textbf{y} \in \mathcal{D}_0} f(\textbf{x} - \textbf{y}) d\textbf{y}},
\quad {\forall} ~ \textbf{x} \in \mathcal{D}_j.
\label{eq:Vj(x)}
\end{equation}
The range of $\frac{1}{\Lambda_j}$ is a subset of the range of $V_j(\textbf{x})$, and can be expressed as
\begin{equation}
\left. V_j(\textbf{x}) \right|_{\min} \le \frac{1}{\Lambda_j} \le \left. V_j(\textbf{x}) \right|_{\max},
\label{eq:Lambda_Vj(x)}
\end{equation}
where 
$\left. V_j(\textbf{x}) \right|_{\min}$ and $\left. V_j(\textbf{x}) \right|_{\max}$ are the minimum value and maximum value of $V_j(\textbf{x})$, respectively.
The properties of $\left. V_j(\textbf{x}) \right|_{\min}$ and $\left. V_j(\textbf{x}) \right|_{\max}$ are summarized in the below two lemmas.

\medskip
\begin{lem}
	\label{lem:minV}
	The properties of $\left. V_j(\textbf{x}) \right|_{\min}$ are listed as follows:
	\begin{itemize}
		\item $ \left. V_j(\textbf{x}) \right|_{\min} \le V_j(\textbf{x}_{O_j})$,
		
		\item The upper bound of $\left. V_j(\textbf{x}) \right|_{\min}$ is
		\begin{equation}
		\left. V_j(\textbf{x}) \right|_{\min}
		\le
		\begin{cases}
		\frac{\frac{2}{3}{R_j^{ - 1.5}}}{d_1^{ - 1.5}\left(\frac{7}{6} + \ln \frac{{{d_1}}}{{{d_0}}}\right)} ,                    &  d_1 < R_j \le 2R_0, \\
		\frac{ \ln \frac{d_1}{R_j} + \frac{2}{3} }{  \ln \frac{d_1}{d_0} + \frac{7}{6}},  & d_0 < R_j \le d_1, \\
		1 - \frac{R_j^2}{2 d_0 \left(\ln \frac{d_1}{d_0} + \frac{7}{6} \right)}, &    0<R_j \le d_0.
		\end{cases}
		\label{eq:Vj_min}
		\end{equation}
	\end{itemize}
\begin{IEEEproof}
	Please see Appendix A.
\end{IEEEproof}
\end{lem}
Lemma \ref{lem:minV} reveals that the upper bound of $\left. V_j(\textbf{x}) \right|_{\min}$ is achieved when $\textbf{x} = \textbf{x}_{O_j}$, which is the center point of $\mathcal{D}_j$. 
By substituting $\textbf{x} = \textbf{x}_{O_j}$ into \eqref{eq:Vj(x)} and with further manipulation,  the upper bound of $\left. V_j(\textbf{x}) \right|_{\min}$ is derived as \eqref{eq:Vj_min} shown.

\medskip

\begin{lem}
\label{lem:maxV}
The properties of $\left. V_j(\textbf{x}) \right|_{\max}$ are listed as follows:
\begin{itemize}
	\item 
	$\left. V_j(\textbf{x}) \right|_{\max} = V_j(\textbf{x}')$, where $\textbf{x}'$ is the coordinates of the point located at the boundary of $\mathcal{D}_j$ and closest to $O_0$.
	
	
	\item The upper bound of $\left. V_j(\textbf{x}) \right|_{\max}$ is
	\begin{equation}
	 \left. V_j(\textbf{x}) \right|_{\max}  \le 1- \frac{ q(R_j)} { {{2 \pi d_1^{-1.5}} \left( \ln \frac{d_1}{d_0} + \frac{7}{6} \right)} },
	 \label{eq: Vj_max}
	 \end{equation}
	where $q(R_j)={\int_{\textbf{y} \in \mathcal{D}_j} f(\textbf{x}' - \textbf{y}) d\textbf{y}}$ is an increasing function of $R_j$,
	and $ 0 < q(R_j) \le {{ \pi d_1^{-1.5}} \left( \ln \frac{d_1}{d_0} + \frac{7}{6} \right)}$. 
	
%
	
	\item  $0 < \left. V_j(\textbf{x}) \right|_{\max} \le 1$.
	
	\item  $ \left. V_j(\textbf{x}) \right|_{\max} \ge \frac{1}{2}$ for $R_j \le R_0/2$.	
\end{itemize}
\begin{IEEEproof}
	Please see Appendix B.
\end{IEEEproof}
\end{lem}
Lemma \ref{lem:maxV} shows that $\left. V_j(\textbf{x}) \right|_{\max}$ is achieved when $\textbf{x} = \textbf{x}'$. 
Through substituting $\textbf{x} = \textbf{x}'$ into  \eqref{eq:Vj(x)} and with further manipulation,
the upper bound of $\left. V_j(\textbf{x}) \right|_{\max}$ can be determined as \eqref{eq: Vj_max} shown.

Since \eqref{eq:Lambda_Vj(x)} is equivalent to
\begin{equation}
\frac{1}{V_j(\textbf{x})|_{\max}} \le \Lambda_j \le \frac{1}{V_j(\textbf{x})|_{\min}} ,
\end{equation}
the range of $C_j^{\infty}$ can be expressed as
\begin{equation}
\log \frac{1}{V_j(\textbf{x})|_{\max}} \le C_j^{\infty} \le \log \frac{1}{V_j(\textbf{x})|_{\min}} ,
\end{equation}
which means 
\begin{equation}
C_j^{\infty}|_{\min} = \log \frac{1}{V_j(\textbf{x})|_{\max}}, ~~ 
C_j^{\infty}|_{\max} = \log \frac{1}{V_j(\textbf{x})|_{\min}}.
\label{eq:Cj_range_Vj}
\end{equation}
Combining \eqref{eq:Cj_range_Vj} with Lemma \ref{lem:minV} and Lemma \ref{lem:maxV}, we get the following theorem for $C_j^{\infty}|_{\min}$ and $C_j^{\infty}|_{\max}$:


\bigskip
\begin{thm}[Average cluster capacity region]
	\label{thm:Cj_LB}
	Conditioning on a round network area with a constant user density, and a given three-slope path-loss model, the lower bound of $\left. {C_j^{\infty}} \right|_{\max}$  for a round cluster can be determined by
	\begin{equation}
	\left. {C_j^{\infty}} \right|_{\max} 
	\ge
	\begin{cases}
	{\log} \left[{{R_j}^{1.5} {d_1}^{-1.5} \left( \frac{3}{2} \ln \frac{d_1}{d_0} + \frac{7}{4} \right)} \right] ,   &  d_1 < R_j \le 2R_0, \\
	{\log} \left({\ln \frac{d_1}{d_0} + \frac{7}{6}}\right)  - {\log}_2 \left({ \ln \frac{d_1}{R_j} + \frac{2}{3} } \right),  & d_0 < R_j \le d_1, \\
	- {\log}_2 \left[1 - \frac{R_j^2}{2 d_0^2 \left(\ln \frac{d_1}{d_0} + \frac{7}{6} \right)}\right], &    0< R_j \le d_0.
	\end{cases}
	\label{eq:Cj_max_LB}
	\end{equation}
	The lower bound of $\left. {C_j^{\infty}} \right|_{\min}$ can be determined by
	\begin{equation}
	\left. C_j^{\infty} \right|_{\min} 
	 \ge \log_2\frac{{2 \pi d_1^{-1.5}} \left( \ln \frac{d_1}{d_0} + \frac{7}{6} \right)}{{2 \pi d_1^{-1.5}} \left( \ln \frac{d_1}{d_0} + \frac{7}{6} \right) - q(R_j)}.
	\label{eq:Cj_min_LB}
	\end{equation}
\end{thm}

\bigskip

It can be observed that both the bounds for $\left. C_j^{\infty} \right|_{\max} $ and $\left. C_j^{\infty} \right|_{\min} $ are increasing functions of the cluster radius $R_j$, consistent with the aforementioned result that $C_j^{\infty}$ and cluster area $\lvert \mathcal{D}_j \rvert$ are positively correlated. The physical insight is that the more nodes in a cluster performing coordinated communications, the larger cluster capacity achieved.
Moreover, by substituting $R_j^*$ for $R_j$ in (\ref{eq:Cj_max_LB}) and   (\ref{eq:Cj_min_LB}), we can obtain the lower bounds for $\left. C_j^{\infty} \right|_{\max} $ and $	\left. C_j^{\infty} \right|_{\min} $ of a C$^2$ architecture.
As such, Theorem \ref{thm:Cj_LB} provides a substantially simplified way to determine the region of average cluster capacity, reducing the complexity from high-dimensional matrix manipulation (shown as  (\ref{eq:C_j_infty})) to pure numerical calculation. 
In practice, Theorem \ref{thm:Cj_LB} can be applied for evaluating real-world networks, whose region can be regarded as round and users are  distributed with a constant density.

\begin{figure}[htbp]
	\centering
	\includegraphics[scale=0.7]{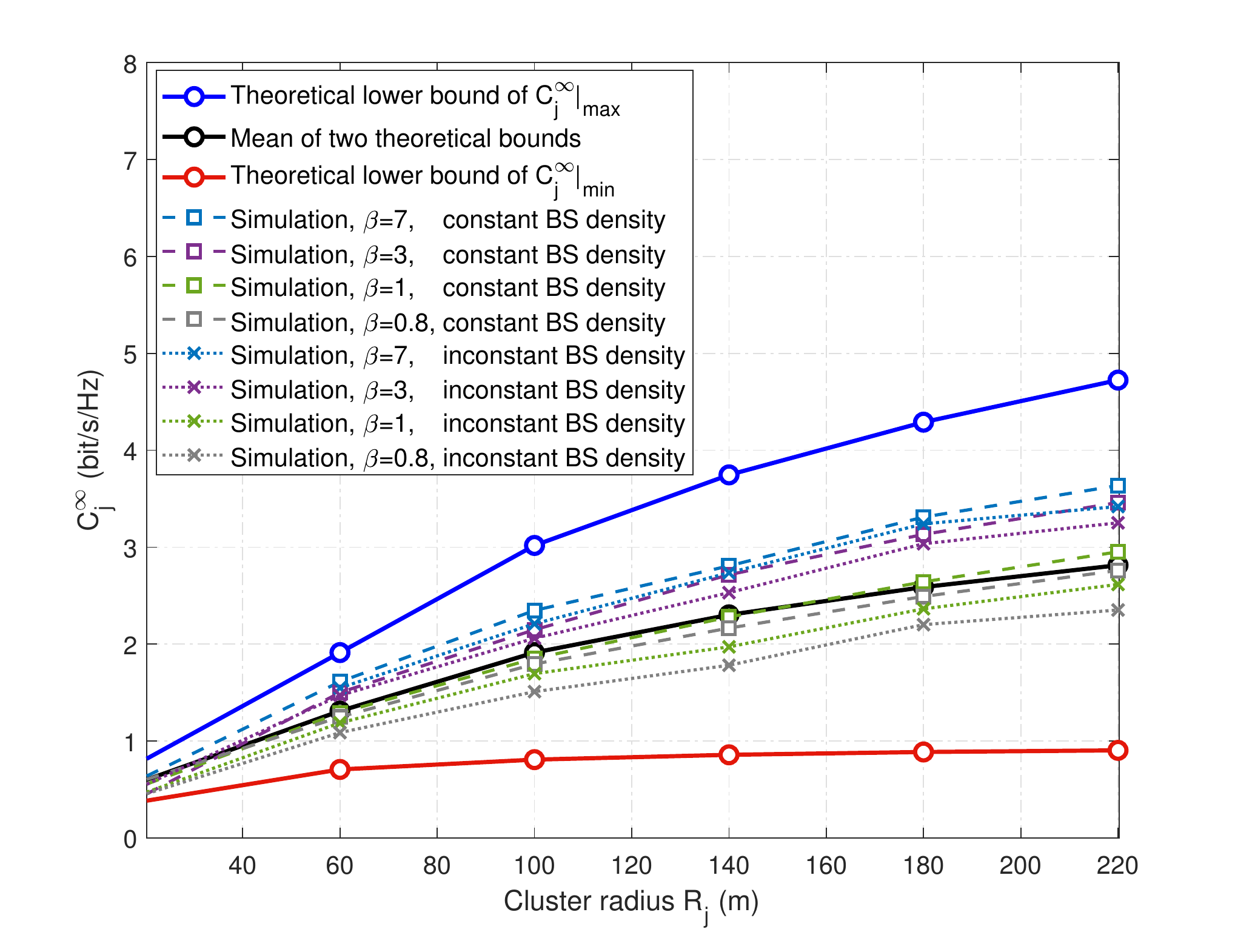}
	\caption{\textbf{Theoretical and simulation results for the average cluster capacity per BS versus cluster radius with a constant user density. } 
		Simulation results (8 dashed lines) contain the scenarios of constant and inconstant BS densities, and four different ratios between the number of users and the number of BSs ($\beta = 0.8, 1, 3, 7$). }
	\label{fig:CjmaxCjminCjave}
\end{figure}

To verify Theorem \ref{thm:Cj_LB}, we plot both of our derived theoretical bounds and  a series of simulated $C_j^{\infty}$ in Fig. \ref{fig:CjmaxCjminCjave}.
We set the network radius as $R_0 = \unit[1000]{m}$, changing the cluster radius $R_j$ from  $\unit[20]{m}$ to $\unit[220]{m}$, and plot the lower bounds of $C_j^{\infty}|_{\max}$ and $C_j^{\infty}|_{\min}$ for each $R_j$ as a blue curve and a red curve, respectively. The arithmetic mean of these two bounds is plotted as a black curve. 
Note that Theorem \ref{thm:Cj_LB} only restricts the user density as a constant, but does not have any constraint on BS distributions. 
We fix the user density as $\rho_u= \unit[6 \times 10^{-3}]{m^{-2}}$, and generate a series of C$^2$ architectures with different parameter settings on BS distributions, to plot the simulated $C_j^{\infty}$ according to its original definition (\ref{eq:C_j_infty}). 
The parameter settings include four values of $\beta$ and two kinds of BS densities, one is constant and the other is centrally dense and peripherally sparse. 
It can be observed that the theoretical bounds of $C_j^{\infty}|_{\max}$ and $C_j^{\infty}|_{\min}$ indeed bound the range of simulated $C_j^{\infty}$ with various parameter settings, and thus can be directly utilized to evaluate the network performance in practice. Moreover, the simulated $C_j^{\infty}$ is positively correlated to the ratio $\beta$ for a given BS distribution; $C_j^{\infty}$ approaches the bound of $C_j^{\infty}|_{\max}$ as $\beta$ increases, while approaches the bound of $C_j^{\infty}|_{\min }$ as $\beta$ decreases.
This is because the larger $\beta$ corresponds to the fewer BSs, and $C_j^{\infty}$ is inversely proportional to the number of BSs as defined in  (\ref{eq:C_j_infty}). Another observation that we wish to point out is that, the simulated $C_j^{\infty}$ approaches the black curve as $\beta$ approaches $1$.
This implies that the average value of the two theoretical bounds can be regarded as an approximation on $C_j^{\infty}$ for real-world networks with similar numbers of BSs and users.

\section{{Performance comparison across C$^2$, CoMP, and cellular architectures}}
In this section, we will compare the performance of three network architectures, including our proposed C$^2$ architecture and two conventional BS-centric architectures: CoMP and cellular.

In simulation, we set up a round network region with radius $R_0 = \unit[1000]{m}$, where we generate users and BSs with location coordinates randomly chosen from continuous uniform distributions, with densities of $\rho_u = \unit [6 \times 10^{-3}]{m^{-2}} $ and $\rho_b = \rho_u/\beta$, respectively. Parameters of the near-field boundary and the far-field boundary are chosen as $d_0 = \unit[10]{m}$ and $d_1= \unit[50]{m}$, respectively. The way to emulate three different architectures are briefly introduced below:
\begin{itemize}
	\item The cellular architecture is BS-centric and emulated through each user attaching to its nearest BS, forming the Voronoi tessellation  \cite{voronoi}. 
	\item The CoMP architecture is an enhanced version of the cellular architecture, and is BS-centric as well \cite{DaiBai17}. We emulate it through choosing a datum BS randomly, assembling surrounding BSs within distance $R_{CoMP}$ to this datum as a cluster. Then we choose another datum BS randomly in the remaining BSs and repeat the clustering steps until no BSs remain. Each user attaches to its nearest BS and thus belongs to the cluster this BS is located in.
	\item The C$^2$ architecture can be plotted directly based on our derived optimal cluster number $M^*$ and optimal cluster radius $R_j^*$ with a given cluster radius threshold $R_{Th}$\footnote{Clusters are arranged according to the circle packing in a circle. There may be a few nodes not belonging to any cluster. For these nodes, we perform a fine-grained tuning in simulation by assigning each node to the cluster with minimum distance between the cluster center and this node.}.
	For fair comparisons, $R_{CoMP}$ in CoMP is chosen to be the same as $R_j^*$ in C$^2$. 
	We name the coverage map of the C$^2$ architecture `C$^2$ tessellation'. 
\end{itemize}

\begin{figure}[htbp]
	\centering
	\subfigure[]
	{\includegraphics[scale=0.7]{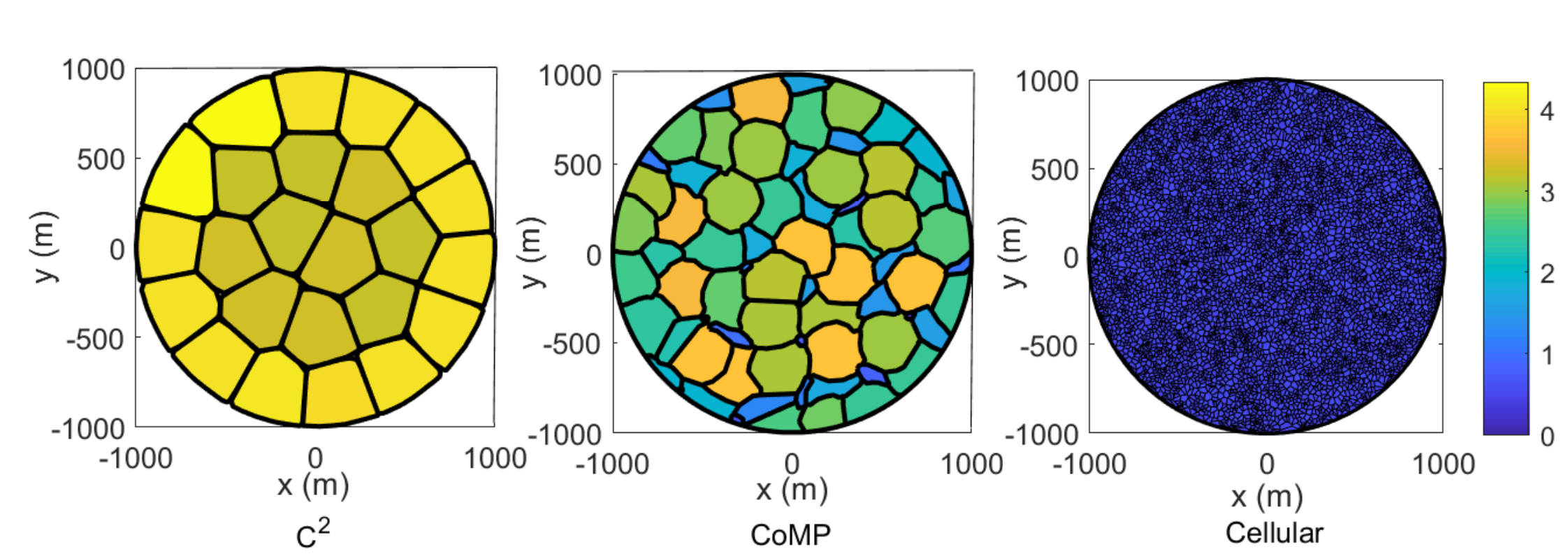}
		\label{fig:heatmap_uniform}} 
	\subfigure[]
	{\includegraphics[scale=0.7]{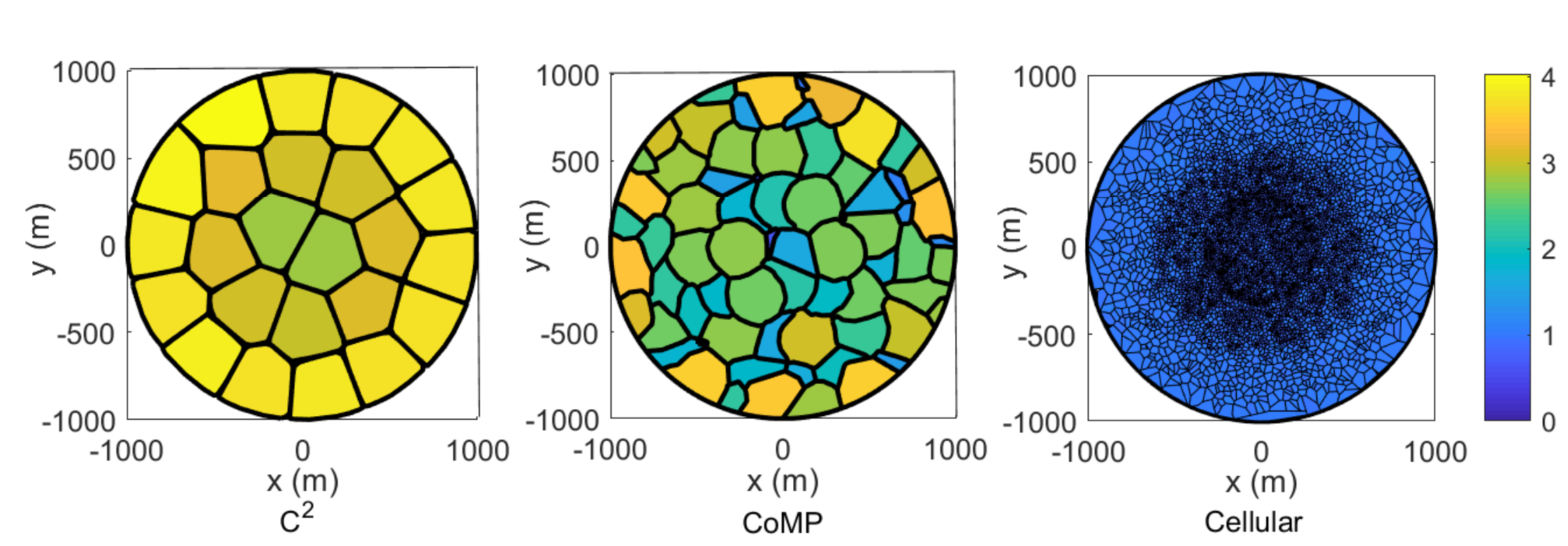}
		\label{fig:heatmap_nonuniform}}
	\caption{\textbf{Heatmaps of the average cluster capacity per BS of every cluster in C$^2$, CoMP, and cellular architectures.} 
		The region circled with a black boundary is a cluster. 
		$R_0= \unit[1000]{m}$, $\beta=3$, $R_{Th}= \unit[175]{m}$.
		Users are distributed with a constant density $\rho_u = \unit[6 \times 10^{-3}]{m^{-2}}$. Two types of BS distributions are considered. 
		\textbf{(a)} BSs are distributed with a constant density.
		\textbf{(b)} BSs are distributed with an inconstant density, which is centrally dense and peripherally sparse.}
	\label{fig:heatmap}
\end{figure}

In Fig. \ref{fig:heatmap}, we plot the heatmaps of  $C_j^{\infty}$ of every cluster in C$^2$, CoMP, and cellular architectures to compare their performance.
Users are distributed with a constant density $\rho_u = \unit[6 \times 10^{-3}]{m^{-2}}$. 
Two types of BS distributions are considered, including the constant density  and the inconstant density, shown in Fig. \ref{fig:heatmap_uniform} and Fig. \ref{fig:heatmap_nonuniform}, respectively.
The inconstant density case refers to BSs deployed centrally dense and peripherally sparse, and please refer to Section IV-A for details. 
It can be observed that the C$^2$ architecture is independent of  BS distributions:  the optimal number of clusters $M^*=24$, and the size of each cluster is the same under different BS distributions. 
In addition, the C$^2$ architecture has the highest $C_j^{\infty}$ compared to its two alternatives, guaranteeing that each BS/user has the highest communication quality.
Different from C$^2$, 
CoMP and cellular are BS-centric architectures, and thus exhibit different forms under different BS distributions in Fig. \ref{fig:heatmap}. Compared to C$^2$, CoMP has a larger difference in cluster size and $C_j^{\infty}$, which implies that it can not guarantee a sufficient communication quality everywhere.  
As for the cellular architecture,  each cluster corresponds to the coverage area of a single BS. Its cluster area is the smallest, and its $C_j^{\infty}$ is the lowest.
As we previously analyzed, cellular can be regarded as an extreme case of our C$^2$ architecture, with the number of clusters taking its maximum value $L$. 
As a brief summary, the C$^2$ architecture provides the highest average cluster capacity per BS/user, and is also the most robust one. Even though the BS distribution is changed, the clusters under the C$^2$ architecture do not change.

\begin{figure}[htbp]
	\centering
	\subfigure[]
	{\includegraphics[scale=0.7]{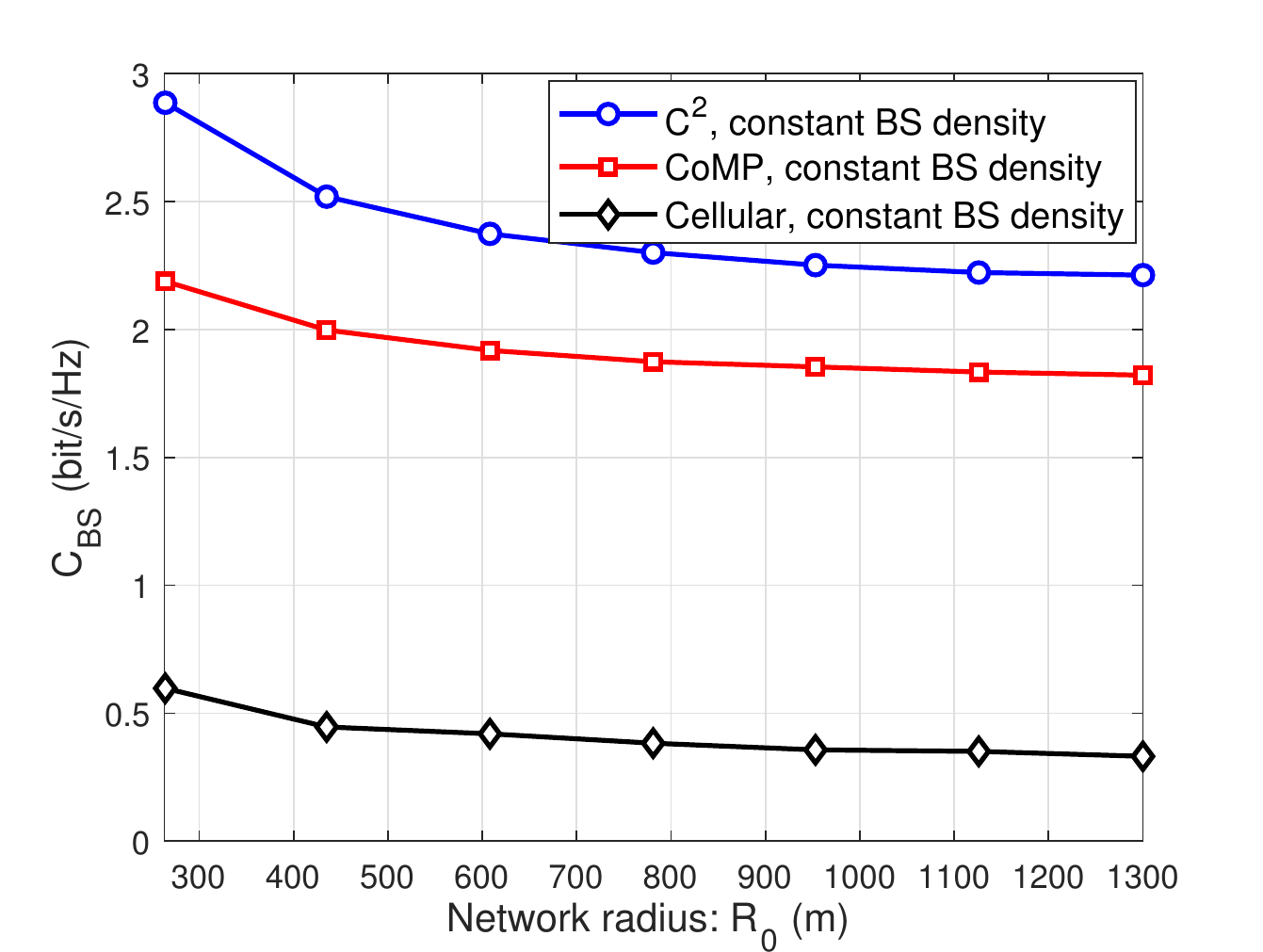}
		\label{fig:performance comparison_uniform_Rth100}} 	
	
	\subfigure[]
	{\includegraphics[scale=0.7]{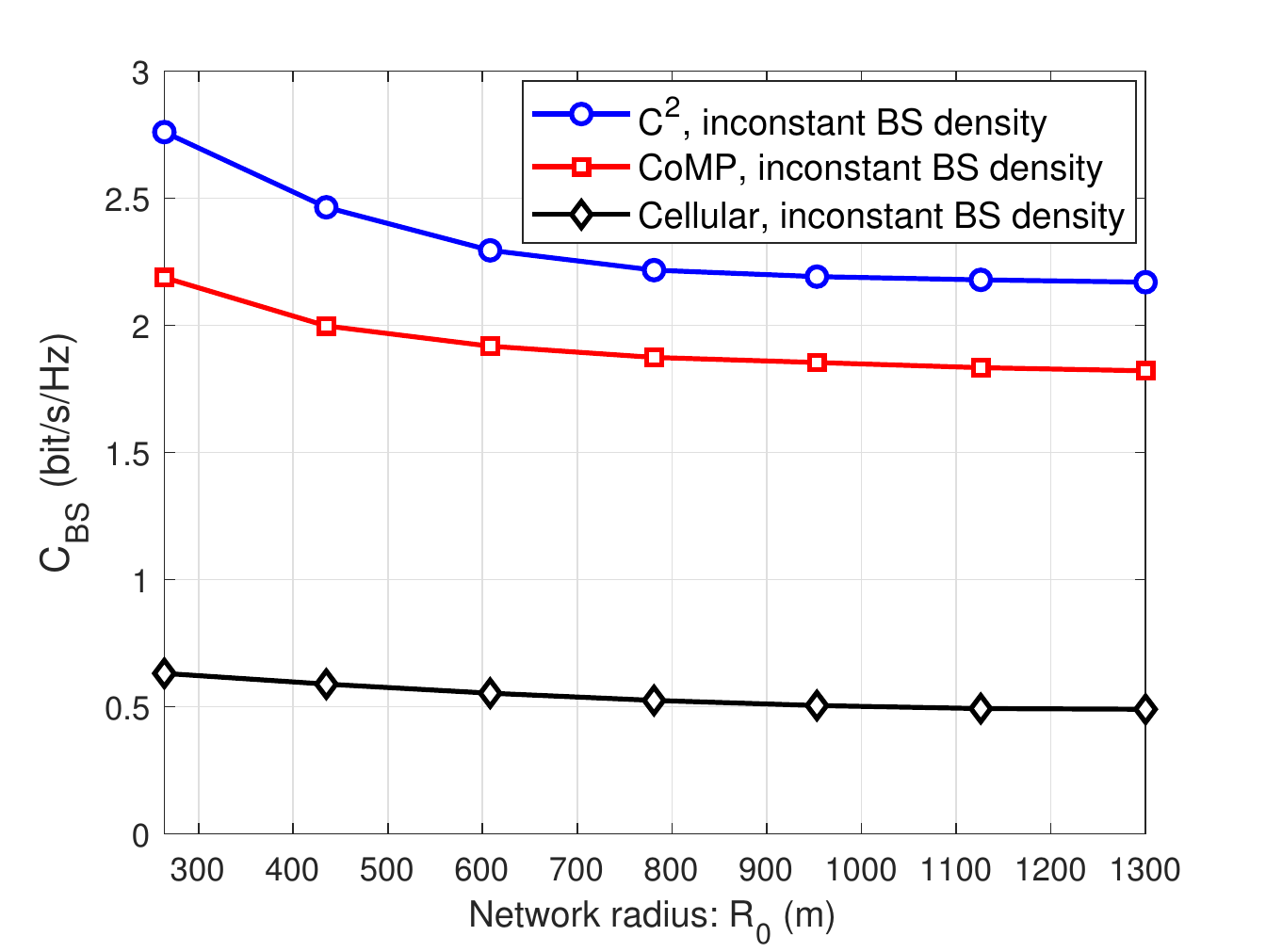}
		\label{fig:performance comparison_nonuniform_Rth100}}
	
	\caption{\textbf{Comparison across C$^2$, CoMP, and cellular architectures
		 in terms of the overall network average capacity per BS versus network radius.} 
		 $\beta=2$, $R_{Th} = \unit[100]{m}$.
		  Users are distributed with a constant density $\rho_u = \unit [6 \times 10^{-3}]{m^{-2}}$.
		  Two different BS distributions are considered. 
	     \textbf{(a)} BSs are distributed with a constant density.
	     \textbf{(b)} BSs are distributed with an inconstant density, which is centrally dense and peripherally sparse. }
	\label{fig:performance comparison_fixedRth}
\end{figure}

After evaluating the cluster performance, we take a step further to evaluate the performance of the overall network. The overall network average capacity per BS, denoted by $C_{BS}$, can be derived based on the average cluster capacity per BS $C_j^{\infty}$ as 
\begin{equation}
C_{BS}  = \frac{\sum_{j=1}^M  C_j^{\infty} L_j}{L}. 
\label{eq:C_per_BS_per_user}
\end{equation}
The overall network average capacity per user is $C_u = \frac{C_{BS}}{\beta}$, only with a difference of constant $\beta$ compared to $C_{BS}$. Here, we choose $C_{BS}$ as the performance metric without losing the generality, since $C_u$ and $C_{BS}$ have the same trends. A comparison across three network architectures in terms of $C_{BS}$ versus the network radius $R_0$ is shown in Fig. \ref{fig:performance comparison_fixedRth}. Our simulation results show that in the case of constant BS density, C$^2$ outperforms CoMP and cellular with gains of at least $21.2\%$ and $383\%$, respectively. In the case of inconstant BS density, which is centrally dense and peripherally sparse (please refer to Section IV-A for details), C$^2$ outperforms CoMP and cellular with gains of at least $17.5\%$ and $315\%$, respectively.

\section{Conclusion}
In this paper, we proposed a C$^2$ architecture for future wireless communication networks. 
It guarantees high capacity for each BS and each user, and exhibits superior scalability with minimal signaling overhead simultaneously. 
The C$^2$ architecture is designed based on our average cluster capacity theorem, which eliminates high-dimensional matrix calculations and is adaptive to different networks.
The C$^2$ architecture
has excellent generality; both the commercialized cellular architecture and the idealized fully coordinated architecture are its extreme cases. 
Simulation results showed that the C$^2$ architecture has the highest overall network average capacity per BS (or per user).
It outperforms cellular and CoMP architectures with performance gains of at least $300\%$ and $17.5\%$, respectively. 
The C$^2$ architecture has excellent robustness, in the sense that its clusters will not change even if the BS distribution is changed.
Last but not the least, we found that the BS distribution should not be regarded as the core factor in future network designs, since it will not affect the C$^2$ architecture with a fixed user distribution. On the contrary, user distributions dominate the average cluster capacity and deserves the focus on next-generation wireless network designs.
All of these findings can be applied in a wide variety of wireless scenarios for architecture designs, such as Wi-Fi, vehicle to everything (V2X) and Industry 4.0, etc.


%
\appendices


\section{Proof of Lemma \ref{lem:minV}}
Based on the fact that $\left. V_j(\textbf{x}) \right|_{\min}$ is less than or equal to $V_j(\textbf{x}_{O_j})$, we have
	\begin{equation}
		\begin{aligned}
			\left. V_j(\textbf{x}) \right|_{\min}
			& 
			\le V_j(\textbf{x}_{O_j})\\
			& =\frac{ \int_{\textbf{y} \in \mathcal{D}_0 - \mathcal{D}_j} f \left(\textbf{x}_{O_j} - \textbf{y} \right) d\textbf{y}}  { \int_{\textbf{y} \in \mathcal{D}_0} f(\textbf{x}_{O_j} - \textbf{y}) d\textbf{y}}  \\ 
			& \le \frac{\left. {\int_{\textbf{y} \in \mathcal{D}_0 - \mathcal{D}_j} f(\textbf{x}_{O_j} - \textbf{y}) d\textbf{y}} \right|_{\mathcal{D}_0 \to \infty} }  {\left. \int_{\textbf{y} \in \mathcal{D}_0} f(\textbf{x}_{O_j} - \textbf{y}) d\textbf{y} \right|_{\mathcal{D}_0 \to \infty} }, 
		\end{aligned} 
	\end{equation}
	where the denominator can be calculated as
	\begin{equation}
		{\left. \int_{\textbf{y} \in \mathcal{D}_0} f(\textbf{x}_{O_j} - \textbf{y}) d\textbf{y} \right|_{\mathcal{D}_0 \to \infty} } 
		= 
		{2\pi {\rho _u}d_1^{ - 1.5}\left(\frac{7}{6} + \ln \frac{{{d_1}}}{{{d_0}}}\right)},
	\end{equation}
	and the numerator is derived as
	\begin{equation}
		{\left. {\int_{\textbf{y} \in \mathcal{D}_0 - \mathcal{D}_j} f(\textbf{x}_{O_j} - \textbf{y}) d\textbf{y}} \right|_{\mathcal{D}_0 \to \infty} }
		=
		\begin{cases}
			{\frac{4}{3}\pi {\rho _u}{R_j^{ - 1.5}}},                    &  d_1 < R_j \le 2R_0, \\
			2 \pi \rho_u d_1^{-1.5} \left( \ln \frac{d_1}{R_j} + \frac{2}{3} \right),  & d_0 < R_j \le d_1, \\
			2 \pi \rho_u d_1^{-1.5} \left[ \frac{1}{2} \left( 1 - \frac{R_j^2}{d_0 ^2} \right) + \ln \frac{d_1}{d_0} + \frac{2}{3} \right], &    0 < R_j \le d_0.
		\end{cases}
	\end{equation}
	Thus,
	\begin{equation}
		\left. V_j(\textbf{x}) \right|_{\min}
		\le
		\begin{cases}
			\frac{\frac{2}{3}{R_j^{ - 1.5}}}{d_1^{ - 1.5}\left(\frac{7}{6} + \ln \frac{{{d_1}}}{{{d_0}}}\right)} ,                    &  d_1 < R_j  \le 2R_0 , \\
			\frac{ \ln \frac{d_1}{R_j} + \frac{2}{3} }{  \ln \frac{d_1}{d_0} + \frac{7}{6}},  & d_0 < R_j \le d_1, \\
			1 - \frac{R_j^2}{2 d_0^2 \left(\ln \frac{d_1}{d_0} + \frac{7}{6} \right)}, &    0 < R_j \le d_0.
		\end{cases}
	\end{equation}
	The proof of Lemma \ref{lem:minV} is completed. 
	

%
%
%
%

\bigskip

\section{Proof of Lemma \ref{lem:maxV}}
Denote $F_j(\textbf{x}) = \int_{\textbf{y} \in  \mathcal{D}_j} f(\textbf{x} - \textbf{y})   	   d\textbf{y}$, and $F(\textbf{x})=\int_{\textbf{y} \in \mathcal{D}_0} f(\textbf{x} - \textbf{y}) d\textbf{y} $, so that
\begin{equation}
V_j(\textbf{x})
= 1 - \frac{F_j(\textbf{x})}{F(\textbf{x})}, 
\quad {\forall} ~ \textbf{x} \in \mathcal{D}_j.
\end{equation}
It can be obtained that $\forall~ \textbf{x} \in \mathcal{D}_j$,  $F_j(\textbf{x})$ is a constant for $R_0 \le \frac{d_0}{2}$.
$\left. F_j(\textbf{x}) \right|_{\min} = F_j(\textbf{x}') = \int_{\textbf{y} \in \mathcal{D}_j} f(\textbf{x}' - \textbf{y})   d\textbf{y} $ for $R_0 > \frac{d_0}{2}$,
since $F_j(\textbf{x})$ decreases as the distance between $\textbf{x}$ and $\textbf{x}_{O_j}$ increases.
Moreover, $\forall \textbf{x} \in \mathcal{D}_j$, 
	$ \left. F(\textbf{x}) \right|_{\max}  =  F(\textbf{x}') = \int_{\textbf{y} \in  \mathcal{D}_0} f(\textbf{x}' - \textbf{y}) d\textbf{y}$ 
	since $F(\textbf{x})$ decreases as the distance between $\textbf{x}$ and $\textbf{x}_{O_o}$ increases.
	Thus, $ \left. V_j(\textbf{x}) \right|_{\max} = 1 -\frac{F_j(\textbf{x}')}{F(\textbf{x}')} = V_j(\textbf{x}')$.
	Define
	\begin{equation*}
		q(R_j) = {\int_{\textbf{y} \in \mathcal{D}_j} f(\textbf{x}'-\textbf{y}) d\textbf{y}},
		\label{eq:q(Rj)}
	\end{equation*}
	we have
	\begin{equation*}
		\begin{aligned}
			\left. V_j(\textbf{x}) \right|_{\max}
			& =\frac{ \int_{\textbf{y} \in \mathcal{D}_0 - \mathcal{D}_j} f \left(\textbf{x}' - \textbf{y} \right) d\textbf{y}}  { \int_{\textbf{y} \in \mathcal{D}_0} f(\textbf{x}'-\textbf{y}) d\textbf{y}}  \\ 
			& \le \frac{\left. {\int_{\textbf{y} \in \mathcal{D}_0 - \mathcal{D}_j} f(\textbf{x}' - \textbf{y}) d\textbf{y}} \right|_{\mathcal{D}_0 \to \infty} }  {\left. \int_{\textbf{y} \in \mathcal{D}_0} f(\textbf{x}' - \textbf{y}) d\textbf{y} \right|_{\mathcal{D}_0 \to \infty} } \\
			& = 1- \frac{q(R_j)}{{2 \pi  d_1^{-1.5}} \left( \ln \frac{d_1}{d_0} + \frac{7}{6} \right)}. 
		\end{aligned}
		\label{eq:Vj(x)_max}
	\end{equation*}
	It can be derived that $\frac{d}{dR_j}{q(R_j) }>0$,
	which means
	$ q(R_j)$ is an increasing function of $R_j$. 
	Thus, the 
	lower bound and upper bound of $q(R_j)$ can be determined as $R_j \to 0$ and $R_j \to + \infty$, respectively,
	as
	\begin{equation*}
		\begin{aligned}
			q(R_j) & \ge \lim_{R_j \to 0}{\int_{\textbf{y} \in \mathcal{D}_j} f(\textbf{x}'-\textbf{y}) d\textbf{y}}
			= \lim_{R_j \to 0} \pi R_j^2 \times d_1^{-1.5} d_0^{-2}  = 0, \\
			q(R_j)  
			& \le \lim_{R_j\to\infty}{\int_{\textbf{y} \in \mathcal{D}_j} f(\textbf{x}'-\textbf{y}) d\textbf{y}}
			= { \pi d_1^{-1.5}} \left( \ln \frac{d_1}{d_0} + \frac{7}{6} \right).
		\end{aligned}
		\label{eq:q(Rj)_range}
	\end{equation*} 
	As such,
	\begin{equation*}
		\frac{1}{2} \le 1-\frac{ q(R_j) }{{2 \pi d_1^{-1.5}} \left( \ln \frac{d_1}{d_0} + \frac{7}{6} \right)} \le 1 ,
	\end{equation*}
	which means
	\begin{equation*}
		0 < \left. V_j(\textbf{x}) \right|_{\max}  \le 1,
	\end{equation*}
	Moreover, 
	if $R_0\ge 2R_j$, it can be derived that 
	\begin{equation*}
		\left. V_j(\textbf{x}) \right|_{\max}
		=
		\frac{ \int_{\textbf{y} \in  \mathcal{D}_j} f \left(\textbf{x}' - \textbf{y} \right) d\textbf{y}
			+ \int_{\textbf{y} \in  \mathcal{D}_0 - 2 \mathcal{D}_j } f \left(\textbf{x}' - \textbf{y} \right) d\textbf{y}}  { 2 \int_{\textbf{y} \in \mathcal{D}_j} f(\textbf{x}'-\textbf{y}) d\textbf{y} + 
			\int_{\textbf{y} \in  \mathcal{D}_0 - 2 \mathcal{D}_j } f \left(\textbf{x}' - \textbf{y} \right) d\textbf{y} }
		\ge \frac{q(R_j)}{2q(R_j)}
		=\frac{1}{2}.  
	\end{equation*}
The proof of Lemma \ref{lem:maxV} is completed.

\bigskip
\bibliographystyle{IEEEtran}

%
%
%

\end{document}